\documentstyle[twocolumn,subfig]{mn}
\newcommand{\mincir}{\raise
-2.truept\hbox{\rlap{\hbox{$\sim$}}\raise5.truept
\hbox{$<$}\ }}

\newcommand{\magcir}{\raise
-2.truept\hbox{\rlap{\hbox{$\sim$}}\raise5.truept
\hbox{$>$}\ }}

\newcommand{\minmag}{\raise-2.truept\hbox{\rlap{\hbox{$<$}}\raise
6.truept\hbox
{$>$}\ }}

\newcommand{\be}{\begin{equation}}
\newcommand{\ee}{\end{equation}}
\newcommand{\ba}{\begin{eqnarray}}
\newcommand{\ea}{\end{eqnarray}}
\newcommand{\brr}{\begin{array}}

\newcommand{\err}{\end{array}}
\newcommand{\bc}{\begin{center}}
\newcommand{\ec}{\end{center}}

\newcommand{\hmpc}{\,h^{-1}{\rm Mpc}}
\newcommand{\vel}{\,{\rm km\,s^{-1}}}
\newcommand{\hmsun}{\,h^{-1}{\rm M_{\odot}}}

\input{epsf.sty}
\title{Properties of cluster satellites in hydrodynamical simulations}
\author[Tormen, Moscardini and Yoshida]
{Giuseppe Tormen$^1$, Lauro Moscardini$^2$ and Naoki Yoshida$^{3,4}$\\
$^1$Dipartimento di Astronomia, Universit\`a di Padova,
vicolo dell'Osservatorio 2, I--35122 Padova, Italy.
Email: tormen@pd.astro.it \\
$^2$Dipartimento di Astronomia, Universit\`a di Bologna,
via Ranzani 1, I--40127 Bologna, Italy.
Email: moscardini@bo.astro.it \\
$^3$Harvard-Smithsonian Center for Astrophysics,
    60 Garden Street Cambridge, MA 02138, USA. \\
$^4$Present address: National Astronomical Observatory Japan,
Mitaka, Tokyo 181-8588, Japan.
    Email: naoki@th.nao.ac.jp}

\date{\today}

\begin{document}

\maketitle

\begin{abstract}

We analyze the dynamical and thermal evolution of dark matter and ICM
in hydrodynamical Tree-SPH simulations of galaxy clusters.  Starting
from a sample of 17 high-resolution objects, with virial mass ranging
from $3\times 10^{14}$ to $1.7\times 10^{15} \hmsun$, we follow
the build-up of the systems in dark matter and hot gas through the
repeated merging of satellites along their merging history trees.  We
measure the self-bound mass fraction of subhaloes as a function of
time after the merging, estimate the satellite mean orbital properties
as a function of the mass ratio with the main cluster at merging time,
and study the evolution of their internal velocity dispersion, gas
temperature and entropy as the substructure is disrupted by various
dynamical processes, reaching eventually thermo-dynamic equilibrium in
the gravitational potential of the main cluster.  We model some
relevant properties of subhalo orbits, as the time of the first
pericentric and apocentric passages, and the typical distances and
velocities at the corresponding times.  This survival study can be
used to interpret the dynamics of observed merging clusters; as an
example we apply our results to the system 1E0657-56.  We show that,
in the light of our results, the most likely interpretation of the
data for this cluster points to the merger of a small group with mass
$M\approx 1\times 10^{13} \hmsun$ with a massive cluster with
$M\approx 1.3\times 10^{15} \hmsun$.

\end{abstract}


\section{Introduction}

Clusters of galaxies are the largest virialized structures observed in
the universe.  Mainly composed of dark matter, they are also large
repositories of hot plasma, the intracluster medium (ICM).  This makes
them very bright X-ray sources, mainly through free-free emission, and
enables us to observe them out to reasonably high redshifts.

The local abundance of galaxy clusters and its redshift evolution are
powerful tests to discriminate between different cosmological models.
In fact, in the framework of hierarchical clustering, clusters are
expected to form from the repeated merging of smaller, pre-existing,
units.  Depending on the amount of matter present in the universe -
i.e. on the value of the present matter density parameter $\Omega_{\rm
m}$ - we expect clusters to form typically at different redshifts: the
more matter present in the universe, the later in time clusters are
allowed to assemble.

Observations of the ICM portray the cluster structure and its
dynamical stage: relaxed systems may show features consistent with the
presence of cooling flows, although evidence is still contradictory
(e.g. Lewis, Stocke \& Buote 2002; Ettori et al. 2002), while merging
events are clearly associated to inhomogeneities in the ICM
temperature and morphology (e.g.  Markevitch, Vikhlinin \& Mazzotta
2001; Mazzotta, Fusco-Femiano \& Vikhlinin 2002; Markevitch et
al. 2002).

The latest generation of X-ray telescopes, like Chandra and
XMM/Newton, have revealed a wealth of details in the morphological and
spectral properties of clusters, often suggesting the combined action
of several physical processes - one example for all - the role of
thermal conduction and magnetic fields in producing the observed
structure of cold fronts (Vikhlinin, Markevitch \& Murray 2001).
These detailed observations are also showing a complexity in cluster
structure that was not evident in previous observations, and shows
that galaxy clusters are often the site of violent collisions between
different subunits.

Although the hierarchical clustering model seems to fit this picture
perfectly, there are many uncertainties and open questions on the
detailed build-up of matter. Mentioning only the simplest: how often
do clusters undergo major mergings?  How long does it take for the
accreted substructure to relax and virialize in the new gravitational
potential?  What is the effect of the collision on the internal
structure of sub-clumps, and on the overall observables of the system?
In order to address these questions one needs to recur to numerical
simulations that mimic the formation and evolution of galaxy clusters
in a given cosmological model, and analyze the merging history of the
forming systems as they grow in time.  While a fairly extended
literature has been published on the evolution of the dark matter
component in the assembly of galaxy clusters (Tormen, Diaferio \& Syer
1998; Wechsler et al. 2002; Zhao et al. 2003; Taffoni et al. 2003),
surprisingly little work exists which describes the evolution of the
ICM counterpart in a cosmological framework. The main reason for this
is that any study of the ICM requires performing N-body hydrodynamical
simulations; now, producing a sample of simulated clusters with
sufficient spatial and mass resolution to follow the details of the
ICM evolution is still a demanding task, both for the practical setup
(choosing a cosmological simulation or performing several
re-simulations, as described below) and for the rather heavy
computational requirements.  In fact, while there has been a lot of
investigation of the physics of merging cluster using simulations,
(e.g. Roettiger, Burns \& Loken 1996; Ricker \& Sarazin 2001; Ritchie
\& Thomas 2002), these studies mostly used non-cosmological initial
conditions in order to address selected and specific theoretical and
observational issues, and disentangle them from the convolution of
cosmological merging history.  In this paper we use the somewhat
orthogonal viewpoint of analysing the history and properties of
cluster collisions in a self-consistent cosmological environment.
This approach guarantees self-consistency of the cluster structural
properties (like profiles and shape) and makes it easy to give a
statistical meaning to our results.

An important step towards the possibility of realising such a program
has been taken by the public release of the Tree-SPH code GADGET
(Springel, Yoshida \& White 2001), which has soon established itself
as one of the reference cosmological codes for the simulations of
large-scale structure and of single systems.  In this paper we combine
the use of GADGET and the technique of {\em single system
re-simulation} (described below; see also Tormen, Bouchet \& White
1997) to produce a sample of 17 high-resolution hydrodynamical
clusters.

Whereas in principle simulations should contain as much physics as is
reasonable to model, we decided to start from the simplest possible
case: that of a non-radiative gas.  That means that our simulations
model processes like dissipation and shock heating, but do not include
radiative cooling, nor more complex processes like star formation or
energy feed-back (e.g. Yoshida et al. 2002; Muanwong et al. 2002;
Tornatore et al. 2003).  This choice is justified by several reasons:
first, runs with cooling and extra physics are much more demanding in
terms of computational costs, making it much harder to obtain a sample
as large as ours at the same resolution.  Second, the practical
implementation of a simulation becomes more and more uncertain as
extra physics is added. In particular, as of today no real agreement
has been reached on the effect of, e.g. cooling on the global
properties of clusters nor on the numerical convergence of the results
(e.g. Balogh et al. 2001).  Finally, we deem it necessary to proceed
by steps, and the step immediately next to pure N-body simulation is
the one we have taken of a non-radiative gas.  Proceeding by steps
will hopefully allow a better interpretation and understanding of the
results and should guarantee that all the physical and numerical
effect of the simulation are under control.

The outline of the paper is as follows. In Section~\ref{sec:method} we
present the simulations and procedures used to create the halo
catalogues and merging history trees.  In Section ~\ref{sec:timev} we
report our results on the evolution of orbital and internal properties
of the satellites after merging.  In Section~\ref{sec:models} we
present a few analytical fits modelling the previous results.  In
Section~\ref{sec:application} we apply our findings to a real
observation, while in Section~\ref{sec:summary} we summarize and
present our conclusions.  In the Appendix we give explicit formulae
for the analytic fits presented in the paper, and give some scaling
laws useful to translate our results from rescaled to physical units.

\section{Method}
\label{sec:method}

\subsection{Simulations}

\subsubsection{Parent simulation}

The clusters considered in this paper were obtained using the
technique of re-simulating at higher resolution patches of a
pre-existing cosmological simulation, as described in Tormen et
al. (1997).  In this instance the parent simulation is an N-body run
with $512^3$ particles in a box of side $479\hmpc$ (Yoshida, Sheth \&
Diaferio 2001, see also Jenkins et al. 2001).  It assumes a flat
universe with a present matter density parameter $\Omega_{\rm m} =
0.3$ and a contribution to the density due to the cosmological
constant $\Omega_\Lambda = 0.7$. The value of the Hubble constant (in
units of 100 km/s/Mpc) is $h = 0.7$. The initial conditions correspond
to a cold dark matter power spectrum which is normalized to have at
the present epoch a r.m.s. in sphere with radius of $8\hmpc$ equal to
$\sigma_8 = 0.9$.  The particle mass is $6.8 \times 10^{10} \hmsun$,
meaning that cluster-sized haloes are resolved by several thousand
particles; the gravitational softening is $30h^{-1}$ kpc.  From the
output of this simulation corresponding to the present time we
randomly extracted 10 spherical regions of radius between 5 and
$10\hmpc$, each containing either a cluster-sized dark matter halo or
a pair of haloes.

\subsubsection{Generation of new initial conditions}

For each of these regions we built new initial conditions using a
software package developed by one of us (GT), and named ZIC (for {\em
Zoomed Initial Conditions}).  A summary of the procedure follows.  The
position of the particles in the initial conditions of the run define
a Lagrangian region where the initial density field was resampled
using a higher number of particles - on average $10^6$ dark matter
(DM) high resolution (HR) particles - in order to increase spatial and
mass resolution.  On top of each HR particle we placed a gas particle
with identical velocity; we assumed a baryon density parameter
$\Omega_{\rm b}= 0.03$.  The mass resolution of the resimulation
ranges from $2\times 10^9$ to $6\times 10^9 \hmsun$ per DM
particle.  The gravitational softening is given by a $5h^{-1}$ kpc
cubic spline smoothing for all HR particles.

We drastically reduced the number of particles outside the HR
Lagrangian region by interpolating them onto a spherical grid centred
on the geometrical centre of the HR region.  The angular resolution of
the grid was taken between 3 and 5 degrees, corresponding to $50,000$
to $150,000$ low resolution (LR) particles of varying mass and
gravitational softening.  Extensive testing has shown that this
reduction still guarantees a sufficiently correct description of the
tidal field on large scales for most cases.

The HR distribution in the initial conditions contains all the
fluctuations of the matter power spectrum realization of the original
cosmological run, plus a new and independent realization of high
frequency fluctuations from the same spectrum, in order to extend the
spectrum up to the Nyquist frequency of the HR distribution.

The ZIC package has been extensively tested and used to generate
initial conditions for many resimulations at medium to extremely high
resolution (e.g. Tormen et al. 1997; Springel et al. 2001b; Yoshida et
al. 2001b; Stoher et al. 2002).

\subsubsection{Resimulations}

The initial conditions were evolved using the publicly available code
GADGET (Springel, Yoshida \& White 2001) from a starting redshift
$z_{\rm in} = 35-60$ - depending on the run - to $z=0$; during the
runs we took 51 snapshots equally spaced in $\log(1+z)$, from $z=10$
to $z=0$.  This choice gives a typical time spacing of $dt \approx
0.5$ Gyr between successive outputs for $z < 0.8 $.

\subsection{Halo catalogues}

We adopted the spherical overdensity criterion to define collapsed
structures in the simulations.  For each snapshot of each resimulation
we first estimated the local dark matter density at the position of
each particle, $\varrho_{i;{\rm DM}}$, by calculating the distance
$d_{i,10}$ to the tenth closest neighbour, and assuming
$\varrho_{i;{\rm DM}} \propto d_{i,10}^{-3}$.  We then sorted the
particles in density and took as centre of the first halo the position
of the densest particle.  Around this centre we grew spherical shells
of matter, recording the total (i.e. DM + gas) mean overdensity inside
the sphere as it decreases with increasing radius.  We stop the growth
and cut the halo when the overdensity first crosses the virial value
appropriate for the cosmological model at that redshift; e.g.
$\varrho(<r)/\varrho_{\rm b} = 323$ at $z=0$.  The particles selected
in this way belong to the same halo and are used to compute its virial
properties (mass, radius, etc.).  For the definition of virial
overdensity we adopted the model of Eke, Cole \& Frenk (1996).  We
tagged all halo particles as {\em engaged} in the list of sorted
densities, and selected the centre of the next halo at the position of
the densest available particle.  We continue in this manner until all
particles are screened.  We include in our halo catalogue only haloes
with at least $n=10$ DM particles inside the virial radius.  All other
particles were considered field particles.

Occasionally it may happen that large haloes host small knots of dark
matter which are denser than the main halo itself.  In such cases the
density maximum would not pick the correct position as putative halo
centre.  To prevent this kind of errors, whenever a large halo
(defined as one with more than 10,000 particles) was found, we
obtained a second centre estimate as the converged centre of mass of
spheres of decreasing radius (sometimes called the {\em moving centre}
method), and chose - between the two centres - the one for which the
grown halo was more massive.

Another problem originating during the build-up of the halo catalogue
is when a newly found halo intersects one or more haloes previously
found, meaning that the distance between the centres is less than the
sum of the two virial radii.  We deal with such cases as follows:
whenever a halo $A$ is found, we check if it intersects with any other
haloes, and undo all intersecting haloes with mass $M < M_A$ making
their particles available for the build-up of new haloes.  We then
recalculate the mass of halo $A$ considering also the particles
{\em disengaged} by this procedure.  This procedure ensures that - in
the final catalogue - particles located inside the intersections are
always assigned to the most massive of the intersecting haloes.

At redshift $z=0$ this procedure identified 17 massive haloes,
containing on average $N_V \approx 200,000$ dark matter particles
within their virial radius, and a similar number of gas particles.
The corresponding virial masses are in the range $M_V = 3.1\times
10^{14}$ to $1.7\times 10^{15} \hmsun$.

\subsection{Merging history tree}

For each of these cluster-sized haloes we built a merging history tree
using the halo catalogues at all time outputs - separated by redshift
intervals $dz_i$ - as follows.  Starting with a halo at $z=0$, we
define its progenitors at the previous output $z=dz_1$ to be all
haloes containing at least one particle that by $z=0$ will belong to
the first halo.  We call main progenitor at $z=dz_1$ the one giving
the largest mass contribution to the halo at $z=0$.  We then repeat
the procedure, starting at $z=dz_1$ and considering progenitors at
$z=dz_1 + dz_2$, and so on backward in redshift, always following the
main progenitor halo.  In this merging history tree we term {\em
satellites} the progenitors which - at any time - merge with the main
progenitor.  In practice we started from the progenitor list at a
given simulation output and selected those haloes for which at least
one particles was found in the main progenitor at the next time
output.

The actual {\em merging redshift} $z_{\rm mer}$ of a satellite with
the main progenitor - which we define as the redshift when the
satellites first crosses the virial radius of the main halo - is
somewhere in-between the redshift $z_{\rm last}$ - when the satellite
is still found as an individual halo in the catalogue - and the
redshift of the next simulation snapshot, $z_{\rm last} - dz$.  We
formally assigned $z_{\rm mer}$ to each satellite by choosing a random
value uniformly distributed between these two redshifts.  The
evolution of the satellite orbital and dynamical properties is then
studied as a function of the time $\tau$ elapsed after its merging:
$\tau \equiv t(z) - t(z_{\rm mer})$, following it through the
simulation down to $z=0$.

Hereafter we will present results with times in Gyr, while all other
quantities will be expressed in the appropriate units of $h$ for
easier comparison with analogous simulation results (and especially
with Tormen, Diaferio \& Syer 1998), although remember that $h = 0.7$.

To get a rough idea of the merging timescale, for a galaxy cluster
observed at $z=0$ the following relation between lookback-time and
redshift holds: $t_{lb} = 1, 2, 4, 6$ Gyr correspond to $z = 0.075,
0.16, 0.37, 0.65$ for our cosmological model.

\section{Time evolution of the satellites}
\label{sec:timev}

\subsection{Merging of satellites}
\label{sec:merging}

In hierarchical clustering the formation of cosmic structures begins
at high redshifts with the smallest resolved systems and proceeds to
larger ones by repeated merging events.  In this picture galaxy
clusters are located at the highest level of hierarchy, and gather all
the matter accreted through their merging history.  However, numerical
simulations show that this is only approximately true.  The violence
of mergers, the variety of orbits and the complexity of the tidal
field make the merging history of the collapse a more complex process,
so that in general only a fraction $f_c \equiv m_{\rm cap}/m_V$ of the
satellite mass $m_V$ is actually accreted by the main halo; as a
consequence, the sum of all progenitor masses exceeds the mass of the
final halo by approximately 20 per cent, and the excess mass is
finally redistributed in debris within roughly 2 times the halo virial
radius (Tormen 1998).

\subsubsection{Selection criteria}

In our analysis we will consider - without loss of generality - only
satellites donating at least half of their virial mass to the main
halo (i.e. $f_c>0.5$).  This criterion ensures that satellites
contributing only marginally to the merging history are excluded.

Once we have identified all the satellites that merge onto the main
halo progenitor at all redshifts, we can investigate how each
satellite evolves by following in time the fate of the $n_V$ particles
contained inside its virial radius at the last time output before
merging happens.  That is, we can focus on how substructures evolve in
time inside a cluster.

We applied two more selection criteria:
\begin{itemize}

\item[(i)] we only consider satellites with $n_V \geq 100$ dark matter
particles to minimize effect due to numerical resolution;

\item[(ii)] we limit our study to satellites merged after $z_{\rm last}
=
0.8$.
\end{itemize}

The last selection criterion was applied after we realized that the
orbital properties of the satellites, when expressed as function of
the time $\tau$ after merging, depended somewhat on the merging
redshift for $z_{\rm last} \magcir 0.8$, going in the direction of
smaller dynamical timescales for higher values of $z_{\rm last}$; as a
simple rescaling of $\tau$ did not remove this time dependence, we
decided to exclude those data from our study.

Since in our simulations the average mass of the main halo progenitors
is $M_V(z=0.8) \approx 3 \times 10^{14} \hmsun$, and by the end it
grows to $M_V(z=0) \approx 8 \times 10^{14} \hmsun$, the selection in
$z_{\rm last}$ has the side advantage that our results will always
refer to mergers onto cluster-size haloes, and so aquire a more
physical meaning in any comparison with observations.  In total, there
are roughly 1,200 satellites satisfying the above requirements, on
which we will concentrate for our following investigation.

\subsubsection{Definition of satellite mass}

As a satellite comes close to the main halo, dynamical interaction
within the mutual gravitational field will modify the orbital and
internal properties of the satellites.  In particular, tidal stripping
will start unbinding the external regions of the satellite, as they
gradually get heated to the virial temperature of the larger system.
This becomes more dramatic as the satellite crosses the virial radius
of the main halo at $z_{\rm mer}$.

In order to calculate the satellite's properties after $z_{\rm mer}$
we need thus to refer to the part of the satellite that - at each time
$\tau$ - is still self-bound and resisting against disruption.  We
calculated the self-bound fraction of a satellite at each $z < z_{\rm
mer}$, that is, for $\tau > 0$, by the following iterative procedure:
\begin{itemize}
\item[(i)] we locate the satellite densest part by the moving centre
method and take the resulting position as the satellite centre.  We
then make a list of all satellite particles inside $r_V$, the virial
radius of the satellite at $z_{\rm last}$, assuming that all particles
outside this radius are physically unbound to the satellite;

\item[(ii)] we calculate potential and kinetic energy for each selected
particle; velocities are taken in the reference frame of the ten
densest particles; for gas particles we add the internal energy to the
balance;

\item[(iii)] we remove from the list all particles with positive total
energy;

\item[(iv)] we repeat steps (ii) and (iii) until the self-bound mass
converges to some value $m_{\rm sb}(\tau)$ to better than 10 per cent.
Obviously the relation holds $m_{\rm sb}(\tau) \leq m_V$.  This
procedure returns also the converged position and velocity of the
bound part of the satellite.

\end{itemize}

Each satellite donates both dark matter and gas to the main halo.
However, while the dark component is non-collisional and in principle
can freely stream through the cluster centre, ram pressure and
dissipative effects should slow the gas down and bring it to
hydrostatic equilibrium.  Therefore, as the satellite dives into the
main halo, its ICM is stripped out of the dark matter potential well
and the two components physically separate.  The time it takes for
this to happen depends on the mass of the satellite relative to that
of the main halo.

Now, as long as dark matter and gas are united, the orbital properties
of a satellite should be those of the two components together, while
we should follow dark matter and gas as individual systems after they
physically separate.  To discriminate between the two regimes - united
and separated - at each time $\tau$ we calculate the distance between
the dark matter and gas centres, and assume separate components
whenever this distance exceeds $r_V$, the virial radius of the
satellite at $z_{\rm last}$.  This distinction has the practical
consequence - for example - that the potential and kinetic energy of
each particle is calculated considering dark matter and gas as one
system with a common velocity reference frame before separation, but
as two independent particle systems with different velocity frames
after separation.

\subsection{Survival of satellites}
\label{sec:surv}

With our apparatus up and running, we can start asking interesting
questions on the mean properties of merged satellites, like how long -
on average - a satellite remains gravitationally self-bound before
being destroyed and digested by the main halo, what is the rate of
this mass loss, and so on.  Here and below we will show results for
satellites in four different mass ranges: $m_V / M_V < 0.001$, $0.001
\leq m_V / M_V < 0.01$, $0.01 \leq m_V / M_V < 0.1$ and $ m_V / M_V
\geq 0.1$; these intervals will roughly correspond to mergers of -
respectively - small galaxies, massive galaxies, small groups, massive
groups or small clusters, onto a massive cluster of mass $M_V$.

\begin{figure}
\centering
\epsfxsize=\hsize\epsffile{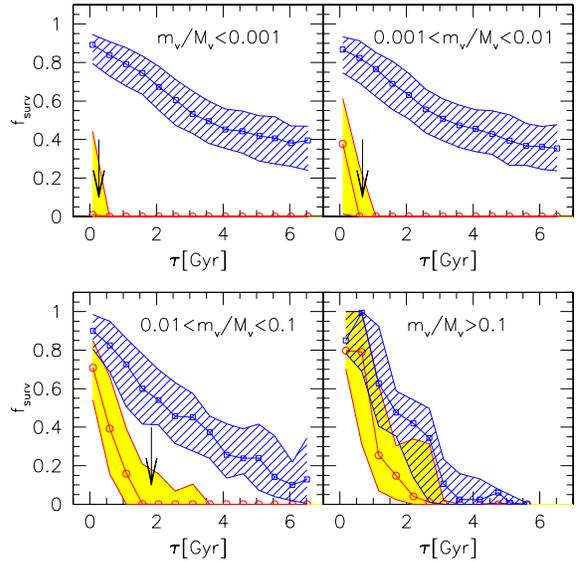}
\caption{Survived mass fraction.  The four panels show the evolution
in time of the self-bound mass of satellites $f_{\rm surv}$ (in units
of their initial donated mass) for satellites in different mass
ranges.  Points indicate median values, bands enclose the central
quartiles of the distributions.  Hatched bands refer to dark matter,
solid ones to gas.  The vertical arrows indicate when, on average,
dark matter and gas separate, as defined in the main text.}
\label{fig:surv_frac}
\end{figure}

In Fig. \ref{fig:surv_frac} we show the evolution of the bound mass
fraction for the dark and gas component, and for each mass range.  To
take into account the fact that only a fraction $f_c$ of the
satellite's mass has actually merged, self-bound fractions $f_{\rm
surv}$ are re-normalized in units of $f_c m_V$.  Dots show the median
values, while the bands enclose the second and third quartiles of the
distributions.

The different behaviour of the two components is evident: dark matter
survives longer in smaller satellites, while the opposite holds for
the gas.  Small, galactic-size objects (upper left panel), maintain a
significant fraction of their initial mass bound to themselves for
many Gyr; on the other hand their ICM is stripped practically as soon
as they cross the virial radius of the main halo.  As we move to
larger masses (upper right and lower left panels) tidal stripping and
dynamical friction become more and more effective in unbinding dark
matter from the satellites; at the same time, the deeper potential
well of the satellites retains a higher fraction of gas for longer
times.  For mergers of comparable masses (lower right panel) dark
matter and gas share a similar fate as they are unbound from the
satellites in 2 to 4 Gyr.

\begin{figure}
\centering
\epsfxsize=\hsize\epsffile{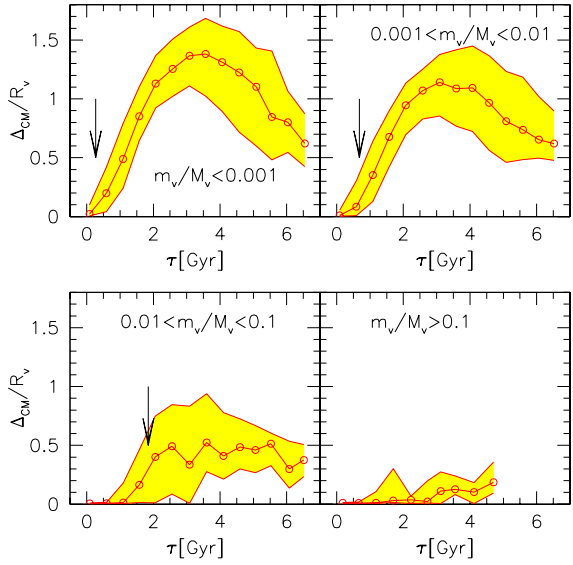}
\caption{Distance between DM and gas satellite centres.  The figure
shows the evolution in time of the separation between the dark matter
and gas components of satellite mass.  Distances are expressed in
units of $R_V$, the virial radius of the main halo at each time
$\tau$.  As before, panels refer to satellites in different mass
ranges.  Vertical arrows indicate the times when the dark matter and
gas components separate.}
\label{fig:delta_CM}
\end{figure}

This behavior is further understood by considering
Fig.\ref{fig:delta_CM}, where we plot the separation $\Delta_{\rm CM}$
between the centre of their dark matter and gas distribution, in units
of the satellite's virial radius, as a function of $\tau$.  The
vertical arrows indicate the average time when the dark matter and gas
components separate - as previously defined.  Median separation times
for the first three mass ranges are at $\tau =$ 0.27, 0.68 and 1.85
Gyr from small to larger masses.  Therefore separation happens sooner
for small lumps and later for more massive ones, while the most
massive satellites (lower right panel) are able to retain their ICM
throughout all their orbit.  In all cases the separation reaches a
maximum value at around $\tau = 3.5$ Gyr, which roughly corresponds -
as we will see below - to the apocentre time for the satellite's
orbit.  The maximum separation is larger for smaller satellites.

Looking back at Fig.\ref{fig:surv_frac}, the same arrows show that, as
soon as the ICM separates from the dark matter component, it becomes
gravitationally unbound.  This is understood as follows: before
separation most of the gravitational potential of the satellite is
provided by its dark matter component, and the gas sits in equilibrium
in its potential well.  After separating from its dark counterpart,
the gas potential well becomes roughly nine times less deep - the
ratio of dark to gas total mass; on the other hand the gas internal
energy is the same as before, that is nine times larger relative to
the new potential: under such conditions the ICM is abruptly unbound
for all satellites.  The most massive satellites are those where the
satellite gas sticks with its dark matter counterpart for longer;
therefore their ICM retains its original properties for a longer time.

\subsection{Orbital properties of satellites}
\label{sec:orbits}

In this section we will investigate the orbital properties of
satellites after they cross the virial radius of the main halo at
$\tau = 0$.  We will consider the evolution of orbital distances and
velocities for all the particles initially associated to a satellite,
and also for the self-bound part of each satellite.  Both cases are of
interest: the former gives indications on how the satellite matter is
redistributed in the new halo, while the latter tells how the core of
a satellite moves through the main cluster.

\subsubsection{Mean orbital properties}

\begin{figure}
\centering
\epsfxsize=\hsize\epsffile{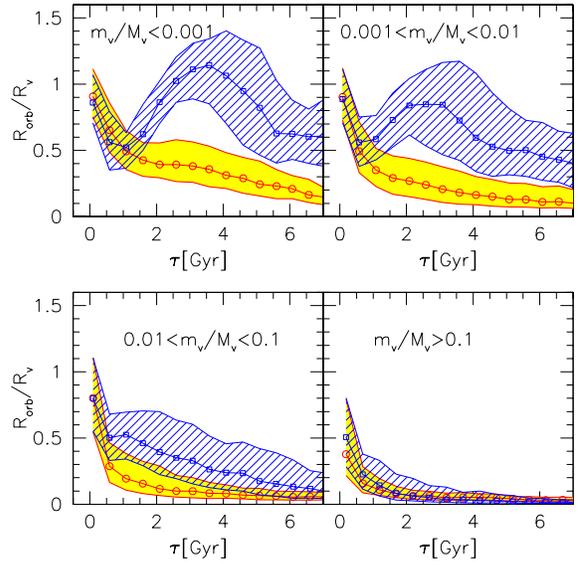}
\caption{Orbital distances $R_{\rm orb}$.
The figure shows the evolution in time of the orbital distance for all
particles in satellites, for different mass ranges.  Symbols, bands
and colours are as in Fig.\ref{fig:surv_frac}.  Distances are in units
of the virial radius $R_V$ of the main halo at each time $\tau$, as in
Fig.\ref{fig:delta_CM}}.
\label{fig:ordist_all}
\end{figure}

\begin{figure}
\centering \epsfxsize=\hsize\epsffile{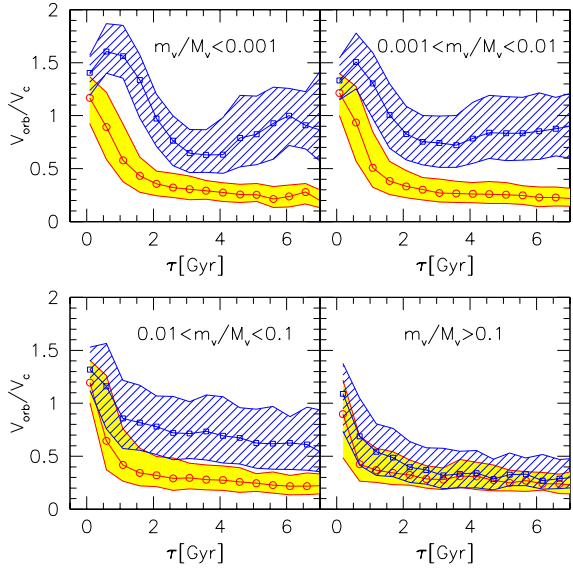}
\caption{Orbital velocities $V_{\rm orb}$.
The figure shows the evolution in time of the orbital velocity for all
particles in satellites, for different mass ranges.  Velocities are in
units of the circular velocity of the main halo: $V_c =
GM_V(\tau)/R_V(\tau)$ at each time $\tau$.  Symbols, bands and colours
are as in Fig.\ref{fig:surv_frac}.}
\label{fig:orvel_all}
\end{figure}

In Figs. \ref{fig:ordist_all} and \ref{fig:orvel_all} we illustrate
the average orbital distance and velocity for all particles belonging
to merged satellites, as a function of $\tau$.  The figures confirm
the different behaviour of the DM and gas particles for small
satellites: while the DM oscillates in position and velocity, the ICM
steadily floats towards smaller and smaller radii, loosing bulk
motion.  Large satellites show instead a similar trend for DM and gas,
as they both slow down and sink towards the centre of the main
cluster.  Maximum (minimum) orbital velocities are reached at
pericentres (apocentres), as expected.

\subsubsection{Orbital properties for the self-bound part}

\begin{figure}
\centering
\epsfxsize=\hsize\epsffile{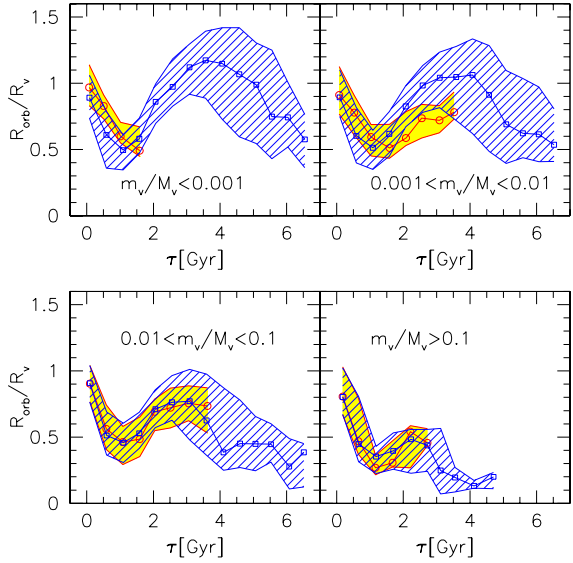}
\caption{Orbital distances $R_{\rm orb}$.
The figure shows the evolution in time of the orbital distance of the
self-bound part of satellites, for different mass ranges.  Symbols,
bands and colours are as in Fig.\ref{fig:surv_frac}.}
\label{fig:ordist_sb}
\end{figure}

\begin{figure}
\centering
\epsfxsize=\hsize\epsffile{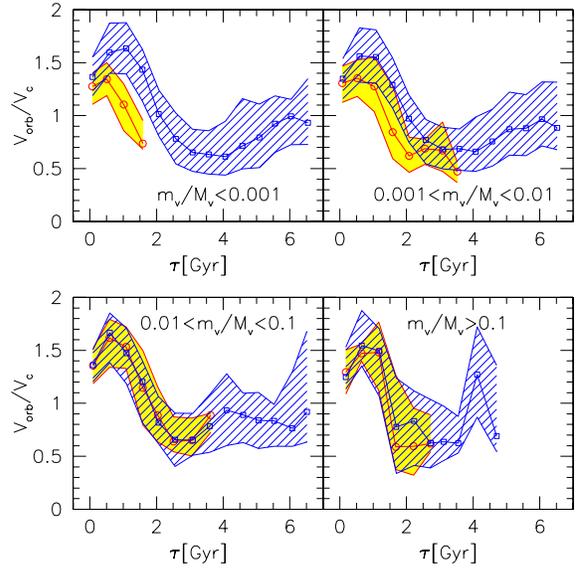}
\caption{Orbital velocities $V_{\rm orb}$.
The figure shows the evolution in time of the orbital velocity for the
self-bound part of satellites, for different mass ranges.  Velocities
are in units of $V_c$, the circular velocity of the main halo at each
time $\tau$.  Symbols, bands and colours are as in
Fig.\ref{fig:surv_frac}.}
\label{fig:orvel_sb}
\end{figure}

The analogue of Figs. \ref{fig:ordist_all} and \ref{fig:orvel_all} for
the satellite self-bound cores is shown in Figs. \ref{fig:ordist_sb}
and \ref{fig:orvel_sb}.  Since the figures reflect the behaviour of
the self-bound satellites, they illustrate the behaviour of the subset
of satellites which - following Fig.  \ref{fig:surv_frac}, have {\em
not} been disrupted by time $\tau$.

The figures show an evolution consistent with the previous results.
We observe that the gas follows the same orbit of the dark component
for longer and longer times in going from small to massive satellites.
In particular, for the smallest satellites (upper left panels of both
figures) the gas slows down and separates dynamically already in less
than 0.5 Gyr, well before reaching the orbital pericentre.  Orbital
velocities increase as satellites fly towards the pericentre, then
decrease toward the apocentre, and so on.  We see that, on average,
satellites merged after $z = 0.8$ have time to complete at most a
couple of orbits; for example, in similar size mergers (lower right
panel) the satellites can reach their pericentre and bounce back, then
finally sink in the core of the main cluster, although by the end they
are still not quite at rest in the cluster potential.

As a summary of the last two subsections, we can remark that the
dynamical behaviour of dark matter and gas in the satellites is
different.  The dark matter component can survive and orbit around the
cluster for quite a long time, while the diffuse baryonic component is
slowed down by the gas already present in the cluster.  For the
largest satellites these differences are reduced, partially because
the dark matter component is quickly dragged toward the cluster centre
by dynamical friction, and also because the potential well of the
satellites is deep enough to retain part of its ICM.

\subsection{Internal properties of satellites}
\label{sec:thermal}

In this section we will address the following issue: how do the
internal properties of a satellite change after it has merged onto the
main cluster progenitor?  The question can be further split in two
considerations: (i) how long does it take for the satellite to {\em
thermalize} into the new environment, that is, how many Gyr are
required before the total dark matter and gas of a satellite are
dispersed and conform to the average internal properties of the main
halo?  And (ii) how do the self-bound part of a satellite react to its
diving into the potential well of the cluster, that is, how will its
velocity dispersion, temperature and entropy change?

\begin{figure}
\centering
\epsfxsize=\hsize\epsffile{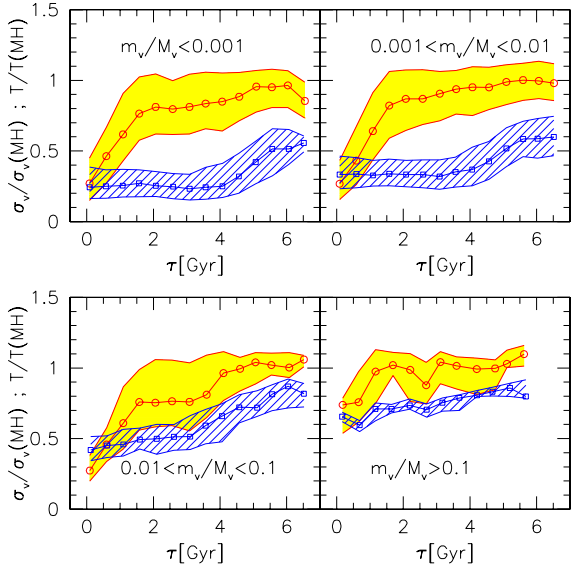}
\caption{Velocity dispersion $\sigma_v$ (hatched bands)
and temperature $T$ (solid bands) for all satellite particles (bound
and unbound).  The values are normalised to the corresponding
quantities for the main halo at each time $\tau$: $\sigma_v({\rm MH})$
and $T({\rm MH})$.  Symbols, bands and colours are as in
Fig.\ref{fig:surv_frac}.}
\label{fig:vel_temp_all}
\end{figure}

\begin{figure}
\centering
\epsfxsize=\hsize\epsffile{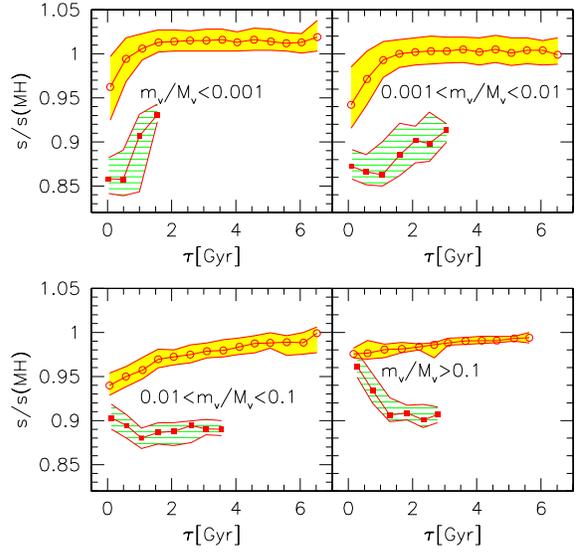}
\caption{Entropy $S$ for all satellites particles (solid band) and for
the self-bound ones (hatched band).  The values are normalised to the
corresponding values for the main halo $S({\rm MH})$ at each time
$\tau$.}
\label{fig:entropy}
\end{figure}

\begin{figure}
\centering
\epsfxsize=\hsize\epsffile{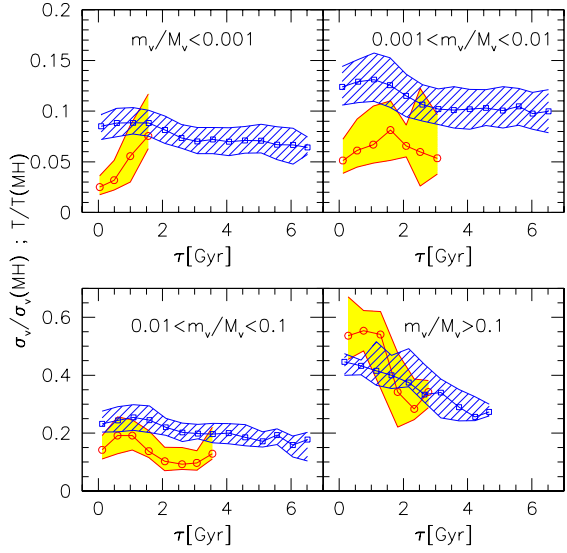}
\caption{Velocity dispersion $\sigma_v$ (hatched bands)
and temperature $T$ (solid bands) for the self-bound satellite
particles.  The values are normalised to the corresponding quantities
for the main halo at each time $\tau$: $\sigma_v({\rm MH})$ and
$T({\rm MH})$.  Symbols, bands and colours are as in
Fig.\ref{fig:surv_frac}.}
\label{fig:vel_temp_sb}
\end{figure}

To answer these questions we looked at the evolution of the dark
matter internal velocity dispersion, and of the gas temperature and
entropy as a function of $\tau$.  Again we did this in two ways: in
Figs.~\ref{fig:vel_temp_all}-~\ref{fig:entropy} we consider all
satellite particles (bound and unbound) in order to answer the first
posed question and see how long it takes for the satellite matter to
reach the virial properties of the hosting halo; in
Figs.~\ref{fig:entropy}-~\ref{fig:vel_temp_sb} instead we consider
only the self-bound satellite particles: this will reveal how the
inner part of a satellite - the one that survives stripping - changes
its properties in reaction to the merging.

\subsubsection{Mean internal properties}

As usual we consider four mass bins according to the mass ratio
between satellite and main halo.  As expected, the satellite's mean
velocity dispersion, temperature and entropy of satellites grows after
merging.  The dark matter velocity dispersion has different behaviour
for small and large satellites.  In the upper panels of
Fig.~\ref{fig:vel_temp_all} (referring to $m_V/M_V < 0.01$),
$\sigma_v$ remains roughly constant for $\tau \approx 4$ Gyr, then it
start growing, without however reaching the virial value.  Larger
satellites (lower panels) show a more gradual growth and reach values
closer to - but still lower than - the mean cluster temperature.
Notice that by the end ($\tau \approx 7$ Gyr) the average $\sigma_v$
of matter in satellites is always {\em lower} than that of the mean
cluster.  This is likely due to the fact that the satellite's internal
velocity dispersion is calculated in a different velocity reference
frame (that of the bound part of the satellite) compared to the
velocity dispersion of the main halo.  Another - more unlikely -
explanation is that the results of Fig.~\ref{fig:vel_temp_all} do not
consider the matter accreted from the field, which may constitute a
non-negligible fraction of the total accreted matter.  However, we
have not verified these hypotheses further.

Considering the temperature, most of the jump in $T$ happens in
roughly $2$ Gyr. After this jump the temperature grows very slowly or
stabilizes.  Satellites in different mass ranges reach different
asymptotic temperatures: compared to the mean temperature $T_V$ of the
main halo, smaller satellites converge to values of $T < T_V$, while
the largest satellites reach $T > T_V$.  This effect is not large, but
it does reflect the final spatial distribution of matter in the main
cluster: small satellites deposit most of their mass in the outskirts
of the cluster because they are stripped of their ICM almost as soon
as they cross the cluster virial radius.  On the other hand, larger
satellites resist stripping and carry part of their mass into the
cluster core.  Since clusters have a larger temperature in the centre
than in the outskirts, the different spatial distribution corresponds
to different gas temperatures.

After an initial jump, entropy (solid band in Fig.~\ref{fig:entropy})
stabilizes in $\tau \approx 2$ Gyr for satellites with $m_V/M_V <
0.01$ (upper panels), while it increases more gradually for larger
satellites.  The asymptotic entropy value decreases going from small to
large satellites, and is slightly higher (lower) than the mean cluster
value for small (large) satellites.  This is again expression of the
overall different spatial distribution of the gas donated by
satellites of different mass, as previously discussed for the
temperature.

>From the overall analysis of
Figs.~\ref{fig:vel_temp_all}-~\ref{fig:entropy} we can conclude that,
while the dark matter component of satellites reaches the virial
equilibrium of the main halo after a time lag that depends on the
satellite mass (determining the dynamical friction time), most of the
gas thermalization process happens in the first $2$ Gyr after merging,
independent of the satellite-to-cluster mass ratio, and of the
different history of the gas (which is stripped from small satellites,
but carried along with larger ones).

\subsubsection{Internal properties for the self-bound part}

Let us now turn to the other question we asked at the beginning of
this section: what is the behaviour of dark matter and gas for that
part of the satellite that remains self-bound?  This issue has
interesting observational consequencies in the discussion of merging
clusters, as the mass of the structures undergoing the merger is
estimated either from the velocity dispersion of their galaxies or
from their X-ray temperature.  It is not a priori guaranteed that
these quantities are indeed a measure of the satellite mass in such
critical dynamical conditions.  The analogue of
Figs.~\ref{fig:vel_temp_all}-~\ref{fig:entropy} for the self-bound
part of each satellite is shown in Figs.~\ref{fig:entropy} and
\ref{fig:vel_temp_sb}.  Here the behaviour is strikingly different
from the one described for
Figs.~\ref{fig:vel_temp_all}-~\ref{fig:entropy}.  In particular we
note that now (i) the starting values of velocity dispersion,
temperature and entropy are usually lower than those previously seen;
(ii) the evolution of these quantities is also different: they not
always grow in time, but often decrease, especially for massive
satellites.

We wish to consider these two points in more detail.  Assuming the
standard scaling laws for velocity dispersion and temperature,
$\sigma_v \propto M^{1/3}$ and $T \propto M^{2/3}$, we expect that
satellites with mass ratio $m_V/M_V = 10^{-4}$, $10^{-3}$, $10^{-2}$
and $10^{-1}$ have velocity dispersion - in units of that of the main
cluster - of the order of $\sigma_v/\sigma_v({\rm main}) = 0.046$,
$0.1$, $0.215$ and $0.464$ and temperature $T_V/T_V({\rm main}) =
0.002$, $0.01$, $0.0464$ and $0.215$.  We stress, however, that these
are only rough estimates.

Now, as for point (i), we can see that the values of the velocity
dispersion for the self-bound part of satellites
(Fig.~\ref{fig:vel_temp_sb}, hatched bands) at merging time, $\tau
\approx 0$, are indeed consistent with those deduced from the scaling
laws.  That means that the self-bound satellites are themselves in
virial equilibrium at merging time, at least as far as their dark
matter component is concerned.  The corresponding ICM temperatures
(solid bands) are instead systematically hotter than their dark matter
counterpart, being a few times larger than the virial estimates.  The
effect is stronger for smaller satellites.  This is telling that the
internal ICM structure of satellites is modified by the collision with
the main halo already at orbiting distances of the order of the virial
radius of the main halo.  We stress that this is visible even in the
self-bound part of the satellite, which is its central and more
shielded part.

Let us compare these results to those of Fig.~\ref{fig:vel_temp_all}
(mean velocity dispersions and temperatures for all satellite
particles: bound and unbound).  There at merging time ($\tau \approx
0$) both the dark matter and gas components were strongly influenced
and heated by the main halo.  Mean temperatures in particular were far
larger than the virial values, except for the most massive satellites
($m_V/M_V > 0.1$, lower right panel).  The gas behaviour is understood
by recalling that small satellites are stripped of their gas content
almost immediately after crossing the virial radius of the main halo,
as shown in the top panels of Fig.~\ref{fig:surv_frac}.

However, velocity dispersions too were higher than expected from
virial relations, although Fig.~\ref{fig:surv_frac} indicates that at
merging time the average self-bound fraction of satellite matter is 80
to 90 per cent.  Evidently the missing 10-20 per cent has been heated
already to the point of contributing significantly to rising the mean
values of Fig.~\ref{fig:vel_temp_all} compared to
Fig.~\ref{fig:vel_temp_sb}.  Entropy shows a similar behaviour:
initial values are higher (Fig.~\ref{fig:entropy}, hatched band).

Passing on to point (ii), that is the evolution of these quantities in
time, we see that the velocity dispersion of the bound part of a
satellite decreases in time.  This means that, while globally the
satellite is heated, its core actually cools down.  Looking at the
temperature evolution, we find the same behaviour for the most massive
satellites (lower panels of Fig.~\ref{fig:vel_temp_sb}). The effect is
particularly dramatic for the mass range $m_V/M_V > 0.1$, where the
ICM cools down by almost a factor two; unlike them, smaller satellites
are heated also in their self-bound part, though less than they are
heated globally.  A similar result is finally found for the entropy:
it increases in small satellites and decreases in large ones, but it
always remains at values lower than the average.

We stress that this result is quite interesting in view of a
comparison with observations; it shows that mass estimate of
substructures undergoing merging cannot be directly derived from,
e.g., their X-ray properties or (assuming that mass traces light) from
the velocity dispersion of their galaxies.  In
Section~\ref{sec:application} of the paper we will apply these results
to observations of a real merging cluster.

\section{Modelling the results}
\label{sec:models}

In this section we come back to the correlation noticed in the mean
behaviour of some of the results presented so far; we will try to
model these correlations parametrically, in order to obtain simple
analytical fits that could be used to improve the dynamical
description of cluster satellites e.g. in semi-analytic models of
galaxy formation.

\subsection{Survived fraction}

In Section \ref{sec:surv} we have investigated the times it takes to
unbind dark and gaseous matter from satellites and shed it in the main
cluster.  Fig.~\ref{fig:surv_frac} shows a typical exponential decay
of the self-bound matter fraction, and it would be nice to be able to
model it in more detail.  To this end we tried the following form:
\be
f_{surv}(\tau) \equiv \frac{m_{\rm sb}(\tau)}{f_c m_V}
= \exp{\left(-\tau/\tau_{\rm dec}\right)} \ ,
\label{eq:fsurv_model}
\ee
where we introduce a {\em decay time} $\tau_{\rm dec}$ as the only
free parameter of this model.

\begin{figure}
\centering
\epsfxsize=\hsize\epsffile{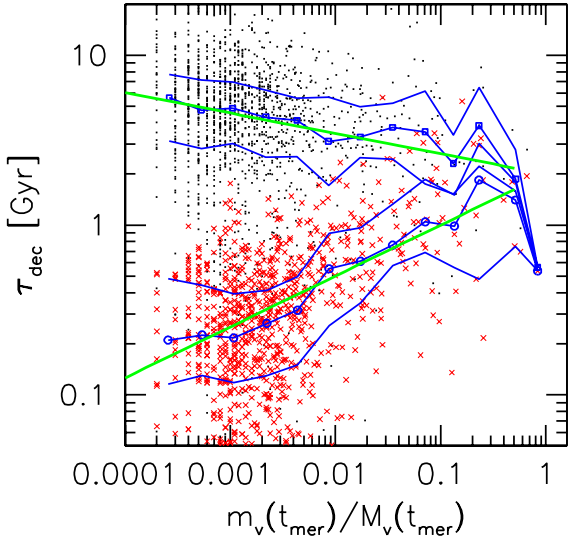}
\caption{Decaying time $\tau_{\rm dec}$ for survived fraction.
We show the log-log plot of the best fitting value of the decaying
time $\tau_{\rm dec}$ for the survived fraction of mass of each
satellite.  Black dots are for DM satellites, red crosses for gas
satellites.  Squares indicate the median value for dark matter as a
function of $m_V/M_V$; circles indicate median values for the ICM
component.  Bands enclose the two central quartiles of the
distributions.  Straight lines are linear fits to the median values.
Their slopes and zero points are given in the main text.}
\label{fig:model_fsurv}
\end{figure}

We applied a $\chi^2$ fit to the curves $f_{\rm surv}(\tau)$ of each
satellite, and obtained the individual best-fitting value of
$\tau_{\rm dec}$.  We did this separately for the dark matter and ICM
components.  In Fig.~\ref{fig:model_fsurv} we show the log-log plot
for the result, as a function of the satellite mass ratio $m_V/M_V$.
The dots show the best-fit $\tau_{\rm dec}$ for each satellite; the
symbols represent the median values for dark matter (squares) and gas
(circles) and the bands enclose the two central quartiles of the
distributions.  The behaviour of median curves is well fitted by
straight lines (superimposed to the curves and symbols) for all
$m_V/M_V \leq 0.5$, with slope and zero point $(-0.12,0.3)$ and
$(0.3,0.3)$ for dark matter and gas, respectively.  The scatter in the
logarithmic residuals is however quite large, as shown in the figure,
being of the order of $\sigma = 0.4$ for both  DM and gas points.

We can immediately translate the above fit into an expression for
$\tau_{\rm dec}$ (expressed in Gyr) as a function of the mass ratio
$m_V/M_V$:
\be
\tau_{\rm dec}(m_V/M_V) = 2 \left(\frac{m_V}{M_V}\right)^{-0.12}
\ee
for dark matter and
\be
\tau_{\rm dec}(m_V/M_V) =  2 \left(\frac{m_V}{M_V}\right)^{0.3}
\ee
for gas.  These expressions, introduced in the model
Eq.~(\ref{eq:fsurv_model}), give
\ba
f_{\rm surv; DM}(\tau) &=& \exp\left[-\frac{\tau}{2}
\left(\frac{m_V}{M_V}
\right)^{0.12}\right]
\nonumber \\
f_{\rm surv; gas}(\tau) &=& \exp\left[-\frac{\tau}{2}
\left(\frac{m_V}{M_V}
\right)^{-0.3}\right]
\ea
for the dark matter and ICM components, respectively.  The large
scatter quoted above should be considered if one is going to use these
relations.

\subsection{Orbital properties}

It is interesting to look in more detail at the mean satellite orbits.
In particular, a simple study we can do is to consider the mean
properties at the first pericentric passage of the satellite, and at
the subsequent first apocentric passage.  We did this by considering
each satellite in turn, and by identifying the first pericentre as the
first relative minimum in the satellite orbital distance going from
$\tau=0$ on.  Analogously, the first apocentre was identified as the
first relative maximum in orbital distance maximum following in $\tau$
the first pericentre.  We limit this study to the first
peri(apo)centric passages because they are easier to find in an
automated way and also because this guarantees a better data
completeness (as fewer and fewer satellites reach further orbital
stages).

\begin{figure*}
\centering
\epsfxsize=\hsize\epsffile{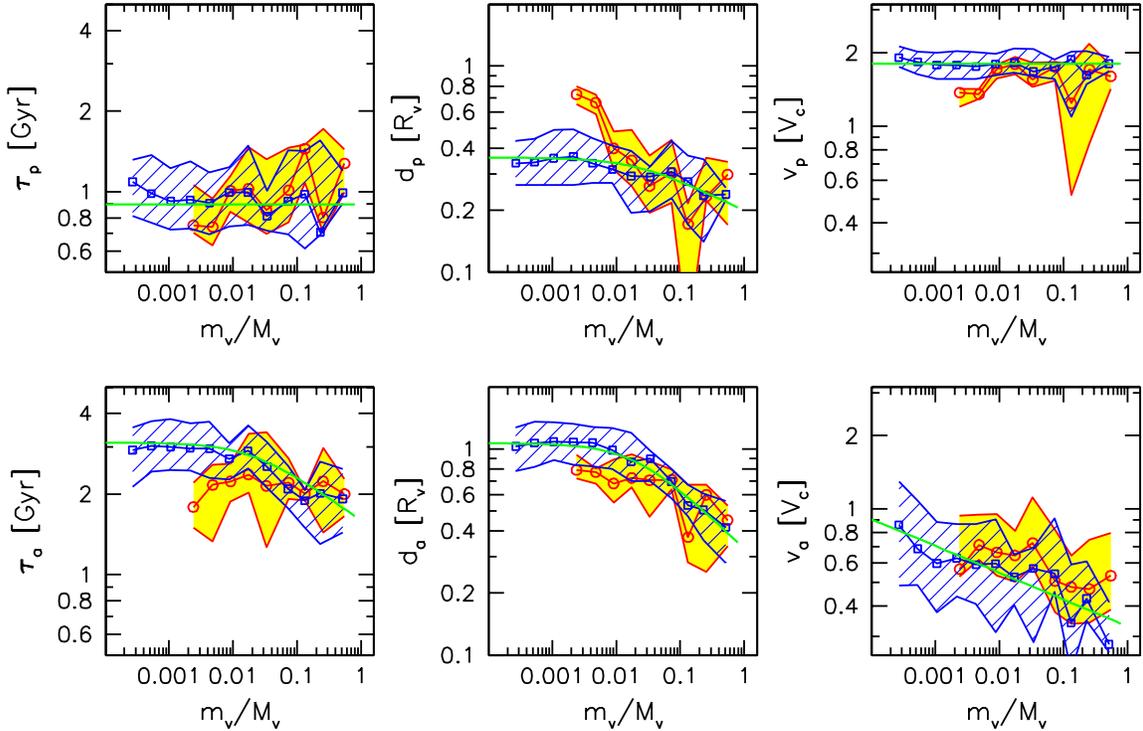}
\caption{The orbital quantities at first
pericentre (upper panels) and first apocentre (lower panels). The time
$\tau$ (in Gyr), the distance $d$ (in units of $R_V$) and the velocity
$v$ (in units of $V_c$) are shown in the left, central and right
panels, respectively. Symbols, bands and colours are as in
Fig.\ref{fig:surv_frac}.  }
\label{fig:model_orbits}
\end{figure*}

For each satellite we then write down the following quantities:
pericentric and apocentric times, distances from the main halo and
orbital velocities.  In Fig.~\ref{fig:model_orbits} we show the median
values and quartile bands for all these quantities, separately for the
dark matter component (square symbols and hatched bands) and for the
ICM (circles and solid bands).

Upper panels refer to pericentre quantities, lower to apocentre ones.
We can note that - on average - satellites reach their pericentre
roughly in 1 Gyr, regardless of their mass.  The dark matter component
reaches a minimum distance of $\approx 0.3 R_V$, with a slight
tendency for large satellites to have more radial and penetrating
orbits; these figures agree with the results found by Tormen (1997),
in the analysis of much smaller N-body simulations of a scale-free
cosmology with $P(k) \propto k^{-1}$.  At pericentre the satellites
reach a maximum speed of the order of twice the circular velocity of
the main cluster.  As already observed in Fig.~\ref{fig:delta_CM}, in
small satellites the ICM is stripped from its dark matter counterpart,
and slows down soon after entering the main halo: in fact gaseous
(solid bands) pericentric distances are larger (and - correspondingly
- velocities are smaller) for $m_V/M_V \mincir 0.01$.

Apocentres for the dark matter component (hatched bands) are reached
in about $\tau \approx 3$ Gyr for small ($m_V/M_V \mincir 0.01$)
satellites, and gradually decrease to $\tau \approx 2$ Gyr for the
most massive satellites (bottom left panel).  The gas components
instead reaches its apocentre in approximately $2$ Gyr, regardless of
the satellite mass.  This again indicates the almost immediate
detachment of the ICM in satellites with $m_V/M_V \mincir 0.01$.

The corresponding distances are larger for small satellites, while
large satellites stay closer to the centre of the main halo (bottom
centre).  Apocentric velocities are a decreasing function of satellite
mass; this might be counterintuitive, as large satellites have smaller
apocentres and one could expect thus larger velocities. Evidently
dynamical friction is extremely efficient in slowing the satellites
down.  The ICM component follows fairly closely the dark matter one,
maintaining only slightly smaller apocentric distances and a hint of
larger apocentric velocities.

The continuous lines superimposed to the median values of the dark
matter component are analytical fits made to model the dark matter
results.  In the Appendix we write down the expression of these fits
for practical use.

\section{Example Application: the Cluster 1E0657-56}
\label{sec:application}


The orbital properties modeled in the previous section can be used as
templates to interpret the dynamics of observed merging systems.  As
an example we choose the cluster 1E0657-56.  This is a merging system
at $z=0.296$, with both X-ray (e.g. Markevitch et al. 2002, 2003, and
references therein) and optical (e.g. Barrena et al. 2002)
observations. Its average X-ray temperature is of the order of $15$
keV, indicating a massive cluster.  The Chandra image shows a
bullet-like substructure off-centre with respect to the cluster core,
probably a galaxy group or small cluster observed in a post-merging
phase.  The temperature of the bullet is of the order of $6-7$ keV,
with a bow-shock feature in front of it.  From the jump in temperature
and density across the shock front Markevitch et al. (2003) derive a
peculiar velocity of $4500^{+1100}_{-800} \vel$ relative to the main
cluster.  Assuming that the merger is almost perpendicular to the line
of sight, from the X-ray bullet-cluster distance of $0.4 \hmpc$ (for
the cosmology considered in this paper) these authors date the
pericentric passage of the bullet at $0.1-0.2$ Gyr ago.

In the optical data of Barrena et al. (2002) the subclump has a
l.o.s. galaxy velocity dispersion of $\approx 210 \vel$, and a l.o.s.
velocity of $\approx 620 \vel$, while the main cluster has a
l.o.s. velocity dispersion of $\approx 1250 \vel$.  The distance
between the core of the main cluster and that of the bullet is 2.7
arcmin, or $0.5 \hmpc$. This value is larger than that derived from
the X-ray data due to an offset between the galaxies and the ICM.  In
particular, galaxies in the subclump are $\approx 0.3 \hmpc$ in front
of the peak of the X-ray emission, as inferred by Fig.1b of Markevitch
et al. (2002).  Assuming for the merger plane a 10 degree deviation
from the plane of the sky, this gives a three-dimensional velocity of
$3500 \vel$, in rough agreement with the derivation of Markevitch et
al. (2002).  Barrena et al. (2002) derive a post-merger mass for the
bullet of the order of $10^{13} \hmsun$, with an uncertainty of a
factor of a few.  They also suggest that the pre-merger mass of the
subcluster could be substantially larger, due to its high X-ray
temperature.  Note that Markevitch et al. (2003) were able to
constrain the dark matter self-interaction cross-section by using four
independents methods, based on the observed gas-dark matter offset,
the absence of an offset between dark matter and galaxies, the high
subcluster velocity and the subcluster survival.

In simulations, dark matter clumps are naturally associated with the
galaxy distribution; the distance between the bullet and the cluster
derived from the optical data should correspond to the distance
between the dark matter structures.  Let us interpret the optical data
in the light of our simulations.  Taking the relations reported in the
Appendix, namely eqs.(\ref{eq:scale1}) and (\ref{eq:scale2}) at face
value, from the measured velocity dispersions the dynamical masses of
the cluster and bullet should be of the order of $1.35 \times 10^{15}
\hmsun$ and $1.04 \times 10^{13} \hmsun$, corresponding to a
mass ratio $1:130$ post merger.

Fig. 8 dates pericentres at $\tau \simeq 1$ Gyr regardless of mass;
taking the Markevitch et al. (2002) datation of 0.1-0.2 Gyr after
pericentre, we derive a time after merging of $\tau \approx 1.1-1.2$
Gyr.  From Fig. 10 we note that the velocity dispersion of the
self-bound part of a subclump (presumably associated with the galaxies
associated to it) does not significantly change in the first 1.5 Gyr
after merging, regardless of its mass.  Therefore the observed ratio
$\sigma_v({\rm bullet})/\sigma_v({\rm main}) \simeq 0.17$ would place
the bullet in a mass intermediate between the second and third panel,
i.e. at around 1:100 mass ratio, also at the pre-merger stage.

This points to a first possible model: that of a galaxy group of
pre-merger mass of the order of $\approx 10^{13} \hmsun$ observed at
$\tau \approx 1.1-1.2$ Gyr, just after its first pericentric passage
through a cluster of mass $\approx 10^{15} \hmpc$.  Fig. 2 shows that
subclumps in the mass range around 1:100 (second and third panels)
still retain - on average - 70 to 80 per cent of their initial dark
matter mass after $\tau \approx 1.1-1.2$ Gyr, while the ICM self-bound
mass may range from 0 to 40 per cent.  The expected velocity after the
first pericentric passage is of the order of $v_p \approx 2V_c \approx
3000 \vel$ (Fig. 8), not inconsistent with the speed inferred by X-ray
data.

The observed separation between the galaxy and X-ray density peaks of
the bullet - estimated from the superimposed X-ray and optical images
- is of the order of $0.11\hmpc$ (proper), or $0.06$ times the virial
radius of the main cluster, which is derived from the cluster mass
and the same scaling laws Eq.(\ref{eq:scale1}) and (\ref{eq:scale2}).
Comparing this value with the one found in simulations (Fig. 3) would
place the mass ratio of the bullet to main cluster more in the range
$1:100$ to $1:10$ (third panel) than in the range $1:1000$ to $1:100$
(second panel).  However, the most puzzling observation is the
temperature of the bullet. From Fig. 10 we see that the expected X-ray
temperature of the self-bound ICM should be at most 20 per cent of the
cluster temperature, or $\approx 3$ keV, while the observations
suggest a ratio of the order of $1:2$.  We can reconcile this
discrepancy if we accept the idea that the X-ray measurements really
refer to the total (bound and unbound) ICM, which from Fig. 8 is shown
to be much higher, at or above the observed 7 keV.

Let us consider the alternative picture by Barrena et al. (2002), of a
more massive subclump of initial velocity dispersion of $700 \vel$,
corresponding to a virial temperature of 7 keV.  In such a case, the
initial mass ratio is of order 0.2, and the post-merger mass would be
of order of 4 per cent.  Now, a subclump this big takes of the order
of 4 Gyr to be disrupted (Fig. 2), and the high speed of the bullet
would imply that we are observing it at its second pericentric passage
(Fig. 5), when the separation between DM and gas is still negligible
(Fig. 3).  However, from Fig. 10 we see that by this time the DM
velocity dispersion is unacceptably high to match the observed $210
\vel$; equally, no ICM has survived to this time, and the X-ray
temperature of the ICM originally in the subclump has reached the
virial value of the cluster (Fig. 8).

In the light of these data we suggest that the most likely
interpretation is the first one, and that the high X-ray temperature
measured for the bullet is due to its ongoing thermalization in the
cluster potential.

\section{Summary and Conclusions}
\label{sec:summary}

This paper studies the properties of the ICM in high-resolution
hydrodynamical simulations of galaxy clusters.  Using a fully
cosmological scenario we focus on the build-up of baryonic and dark
matter through hierarchical clustering. This decomposition allowed us
to see the ICM of the final clusters as the evolution of the ICM of
its progenitors, which undergo a number of dynamical and thermal
changes during and after their merging in the potential well of the
main system. In a way, we tried to answer questions related to the
{\em nature versus nurture} of the ICM.  While the analysis presented in

this paper certainly does not exhaust all questions nor gives all
answers, we think it constitutes a successful attempt to statistically
characterise a few properties of ICM evolution in a cosmological
environment, which may be useful both to understand theoretical issues
and to interpret observational X-ray data.

The main results of the present work may be summarised as follows:
\begin{itemize}

\item
satellites merging onto a cluster undergo a fate that is well
described by one free parameter, namely the mass ratio $m_V/M_V$ of
the satellite to the main object.  The survival time is a decreasing
function of $m_V/M_V$ for the DM component, but an increasing function
of $m_V/M_V$ for the satellite ICM.  For the largest satellites
(almost equal mass ratio) the DM and ICM have similar survival times.

\item
The dark matter and ICM of satellites exhibit quite different
behaviours for different mass ratios.  In small satellites DM and gas
decouple quickly after they cross the virial radius of the main
cluster: while on average the ICM is promptly stripped and unbound by
ram-pressure effects, the DM freely moves and oscillates in the main
halo; this causes a clear spatial separation between the two satellite
components.  In particular, the behaviour of the DM component is quite
consistent with that found by Tormen, Diaferio \& Syer (1998) and
Taffoni et al. (2003), showing that the presence of a diffuse gas has
at most a marginal effect on the global satellite properties.  This
might be expected, as the gravitational contribution of the ICM is of
the order of the cosmological baryonic fraction.

\item
The mean internal properties of satellites (DM velocity dispersion and
ICM temperature) evolve differently from those of the self-bound part:
globally satellites are heated by their encounter with the main halo,
whereas their cores (self-bound parts) can have different behaviours.

On average, the bulk of the ICM initially associated to satellites
thermalizes in the new cluster potential in roughly $2$ Gyr, while the
DM component takes much longer to adapt to the velocity dispersion of
the main cluster.

Considering the self-bound part of satellites, the DM cores always
cool down for all mass ratios, while the core ICM temperature heats up
for small satellites and cools down for massive ones.  This
demonstrates that the galaxy velocity dispersion and the X-ray
temperature of substructure cannot be easily linked to its actual
mass, unless a more thorough analysis of the kinematics and dynamics
of the system is performed.

\item
The ICM entropy shows a trend similar to that of the ICM temperature:
entropy always increases globally, but decreases in the cores of
massive satellites.

\item
We have modeled the mean orbital properties of satellites at their
first pericentre and apocentre, using simple analytic fits.
Applications of these results to optical and X-ray observations of
cluster 1E0657-56 indicates that this system is more likely a merger
between a small group and a massive cluster, observed after its first
pericentric passage.

As a final remark, we remind that our results are based on a set of
adiabatic hydrodynamic simulations.  SPH simulations which include
extra physics, like cooling, star formation and supernova feedback
(e.g. Lewis et al. 2000; Yoshida et al. 2002; Tornatore et al. 2003;
Borgani et al. 2003) show that these processes can modify the global
properties of clusters, and this in turn might affect qualitatively
the results presented here.  However, the present uncertainties on
exactly how to model these processes in simulations, and the possible
influence of other physical mechanisms not yet considered
(e.g. thermal conduction and magnetic fields), makes it impossible to
estimate here the impact of these effects.

\end{itemize}

\section*{Acknowledgments.}

This work has been partially supported by Italian MIUR (Grant 2001,
prot. 2001028932, ``Clusters and groups of galaxies: the interplay of
dark and baryonic matter''), CNR and ASI.  We are grateful to Massimo
Ramella for clarifying discussions.

\appendix
\section{Analytic fits  and scaling laws}

Here we briefly summarize the analytic fits to the results of
Fig.~\ref{fig:model_orbits}, as a function of the mass ratio $x \equiv
m_V/M_V$, where $m_V$ and $M_V$ are the pre-merger virial masses of
the satellite and of the main cluster progenitor, respectively.  In
order to make these quantities physical they must be converted using
the scaling laws also given in Sec.~\ref{section:scalings} below.

\subsection{Analytic fits}

\begin{itemize}
\item
pericentric time:
\be
\tau_p(x) = 0.9 [Gyr]
\ee
\item
pericentric distance:
\be
d_p(x) = 0.2 (x + 0.02)^{-0.15}  [R_V]
\ee
\item
pericentric velocity:
\be
v_p(x) = 1.8  [V_c]
\ee
\item
apocentric time:
\be
\tau_a(x) = 1.6 (x + 0.02)^{-0.17} [Gyr]
\ee
\item
apocentric distance:
\be
d_p(x) = 0.33 (x + 0.02)^{-0.3}  [R_V]
\ee
\item
apocentric velocity:
\be
v_p(x) = 0.33 x^{-0.11}  [V_c]
\ee
\end{itemize}

\subsection{Scaling laws}
\label{section:scalings}

The results presented above are given in rescaled units [distances in
units of $R_V(z)$ and velocities in units of $V_c(z)$] in order to
superimpose results coming from clusters of different mass and at
different redshifts.  In order to apply these models to real
observations, we need to provide suitable conversions to physical or
observational quantities.  We also give the mean conversion relation
between mass and internal velocity dispersion.  The virial masses,
radii, circular velocities and one-dimensional velocity dispersion of
our simulated clusters - at redshift $z \mincir 0.8$ - are well
related by the following relations, which hold for $10^{14} \hmsun
\leq M_V \leq 1.3 \times 10^{15} \hmsun$:

\ba
\left(\frac{V_c(M_V)}{1000 \rm km/s}\right) &=&
1.410 \left(\frac{M_V}{10^{15} {\rm \hmsun}}\right)^{0.353}, \nonumber
\\
\left(\frac{\sigma_v(M_V)}{1000 \rm km/s}\right) &=&
0.982 \left(\frac{M_V}{10^{15} {\rm \hmsun}}\right)^{0.367}, \nonumber
\\
\left(\frac{R_v(M_V)}{\rm \hmpc}\right) &=&
2.164\left(\frac{M_V}{10^{15} {\rm \hmsun}}\right)^{0.294},
\label{eq:scale1}
\ea
where virial radii are in comoving $\hmpc$.  The corresponding
observables in physical coordinates, for systems at redshift $z$, are:
\ba
V_c({\rm phys})      &=& V_c(M_V) \sqrt{1 + z}, \nonumber \\
\sigma_v({\rm phys}) &=& \sigma_v(M_V) \sqrt{1 + z},\nonumber \\
R_v({\rm phys})      &=& R_v(M_V) / (1 + z).
\label{eq:scale2}
\ea

\end{document}